\documentclass[a4paper,fleqn]{cas-sc}

\usepackage[authoryear]{natbib}
\usepackage{xcolor}
\usepackage{booktabs}
\usepackage{multirow}
\usepackage{tablefootnote}
\usepackage{tcolorbox}
\usepackage{listings}
\usepackage{svg}
\usepackage{subcaption}
\usepackage{soulpos}
\usepackage{url}
\usepackage{wrapfig}

\newsavebox\CBox
\def\textBF#1{\sbox\CBox{#1}\resizebox{\wd\CBox}{\ht\CBox}{\textbf{#1}}}

\ulposdef{\myhl}{%
  \mbox{%
    \color{yellow}%
    \rule[-.8ex]{\ulwidth}{11pt}}}

\definecolor[named]{BoxColor}{RGB}{226,233,241}
\definecolor[named]{BoxColorTitle}{RGB}{16,78,139}
\definecolor[named]{lightred}{RGB}{255,222,219}
\definecolor[named]{lightgreen}{RGB}{234,255,233}

\begin{document}

\title [mode = title]{Beyond Questions: Leveraging ColBERT for Keyphrase Search}

\shorttitle{Beyond Questions: Leveraging ColBERT for Keyphrase Search}
\shortauthors{Jorge Gabín et~al.}

\author[lkn,irlab]{Jorge Gabín}[auid=000,bioid=1,orcid=0000-0002-5494-0765]
\cormark[1]
\ead{jorge.gabin@udc.es}
\cortext[cor1]{Corresponding author}
\credit{Conzeptualization, Software, Investigation, Writing - Original Draft, Visualization}

\author[irlab]{Javier Parapar}[auid=000,bioid=2,orcid=0000-0002-5997-8252]
\ead{javier.parapar@udc.es}
\credit{Conzeptualization, Writing - Review \& Editing, Supervision}

\author[uog]{Craig Macdonald}[auid=000,bioid=3,orcid=0000-0003-3143-279X]
\ead{craig.macdonald@glasgow.ac.uk}
\credit{Conzeptualization, Writing - Review \& Editing, Supervision}

\affiliation[lkn]{organization={Linknovate Science},
            city={Santiago de Compostela},
            country={Spain}
}

\affiliation[irlab]{organization={IRLab, CITIC, Computer Science Department, University of A Coruña},
            city={A Coruña},
            country={Spain}
}

\affiliation[uog]{organization={University of Glasgow},
            city={Glasgow},
            country={Scotland}
}

\begin{abstract}
While question-like queries are gaining popularity and search engines' users increasingly adopt them, keyphrase search has traditionally been the cornerstone of web search. This query type is also prevalent in specialised search tasks such as academic or professional search, where experts rely on keyphrases to articulate their information needs. However, current dense retrieval models often fail with keyphrase-like queries, primarily because they are mostly trained on question-like ones. This paper introduces a novel model that employs the \texttt{ColBERT} architecture to enhance document ranking for keyphrase queries. For that, given the lack of large keyphrase-based retrieval datasets, we first explore how Large Language Models can convert question-like queries into keyphrase format. Then, using those keyphrases, we train a keyphrase-based \texttt{ColBERT} ranker (\texttt{ColBERTKP$_{QD}$}) to improve the performance when working with keyphrase queries. Furthermore, to reduce the training costs associated with training the full \texttt{ColBERT} model, we investigate the feasibility of training only a keyphrase query encoder while keeping the document encoder weights static (\texttt{ColBERTKP$_Q$}). We assess our proposals' ranking performance using both automatically generated and manually annotated keyphrases. Our results reveal the potential of the late interaction architecture when working under the keyphrase search scenario. This study's code and generated resources are available at \url{https://github.com/JorgeGabin/ColBERTKP}.
\end{abstract}




\maketitle


\section{Introduction}
\label{sec:intro}
The search process is increasingly adopting question-like queries~\citep{dalton2022conversational,guo2022qanswer,khurana2023natural}. This shift is driven by the enhanced ability of search engines to better process such longer queries and the growing influence of conversational models, which encourage users to phrase their searches as questions. Despite this trend, keyphrase search remains essential, especially in specialised fields such as academic and professional search~\citep{jacso2015academic,li2017investigating,russell2018information,lahiri2024keyphrase}. Experts often use keyphrases and boolean queries to express their information needs. However, nowadays most state-of-the-art neural retrieval models are trained primarily on datasets with question-like queries, such as those from MSMarco~\citep{nguyen2016msmarco}.

\begin{table}[t]
\centering
\renewcommand{\arraystretch}{1.7}
\caption{Highest ranked passages by \texttt{ColBERT} and \texttt{ColBERTKP$_{Q}$} models for a TREC 2019 DL query in both the original and keyphrase format. The ``Label'' column contains the assessment of that document for that query in the qrel file, with – denoting unjudged.} 
\begin{tabular}{llp{11cm}c}
\toprule
\textbf{Doc.} & \textbf{System} & \textbf{Top-1 Passage} & \textbf{Label} \\
\midrule
\multicolumn{4}{c}{\textit{962179: when was the salvation army founded}} \\
\midrule
\rowcolor{lightgreen}
5653659 & \texttt{ColBERT} & The Salvation Army was founded in London's East End in 1865 by one-time Methodist Reform Church minister William Booth and his wife Catherine. Originally, Booth named the organisation the East London Christian Mission. & 3 \\
\rowcolor{lightgreen}
2978866 & \texttt{ColBERTKP$_{Q}$} & The Salvation Army was founded in London's East End in 1865 by one-time Methodist Reform Church minister William Booth and his wife Catherine. Originally, Booth named the organisation the East London Christian Mission. & 3 \\
\midrule
\multicolumn{4}{c}{\textit{962179: salvation army foundation year}} \\
\midrule
\rowcolor{lightred}
5642478 & \texttt{ColBERT} & February 24, 2011. In addition to money, the Salvation Army accepts donations of vehicles, furniture, appliances, clothing and countless other household items. As a religious and charitable organization, the Salvation Army resells your donations to fund its adult rehabilitation programs. Because charitable donations are tax-deductible, the organization provides a receipt as documentation. & - \\
\rowcolor{lightgreen}
2329699 & \texttt{ColBERTKP$_{Q}$} & History. The Salvation Army began in 1865 when William Booth, a London minister, gave up the comfort of his pulpit and decided to take his message into the streets where it would reach the poor, the homeless, the hungry and the destitute. & 3 \\
\bottomrule
\end{tabular}
\label{tab:example}
\end{table}

Here, we focus on the persistent importance of keyphrase queries and propose training dense retrieval models specifically for this query format. Transformer-based~\citep{vaswani2017attention} models like \texttt{BERT}~\citep{devlin2019bert} and, specifically, alternatives like \texttt{DPR}~\citep{karpukhin2020dense}, \texttt{ANCE}~\citep{xiong2021approximate}, \texttt{ColBERT}~\citep{khattab2020colbert}, \texttt{Col$\star$}~\citep{xiao2023reproducibility}, \texttt{GTR}~\citep{ni2022large}, or \texttt{XTR}~\citep{lee2023rethinking}, have demonstrated the effectiveness of dense retrieval models for ranking. While traditional methods use inverted indexes, dense retrieval represents documents and queries as dense vectors, which are then matched using approximate nearest neighbours (ANN) algorithms, such as those present in the \texttt{FAISS} toolkit~\citep{douze2024faiss, johnson2019billion}.

As mentioned above, current dense retrieval models are predominantly trained on question-like queries due to the availability of large-scale datasets covering this specific query type. We argue that this has inadvertently made queries formulated as questions the standard for these models, often overlooking other query formats. Indeed, our observations indicate that traditional dense models might underperform with keyphrase queries. For instance, in Table~\ref{tab:example} we can observe the difference between question-like and keyphrase-like queries. While both models retrieve a highly relevant document for the question-like query, \texttt{ColBERT} struggles with the keyphrase query while one of our keyphrase-tailored models (\texttt{ColBERTKP$_{Q}$}) succeeds.

In this paper, we hypothesise that training those dense retrieval models with a keyphrase-focused dataset may alleviate the problems of current state-of-the-art models when facing this type of query. Keyphrase-tailored models may partially solve the applicability of dense re-ranking in professional, academic, and even web searches, where short keyphrase-like queries continue to be the prevailing query type~\citep{silverstein1999analysis,reimer2023archive,taghavi2012analysis}. Moreover, recognising the computational costs associated with training the full \texttt{ColBERT} model, we investigate the feasibility of training a keyphrase query encoder while keeping the document encoder weights frozen. 

Training neural ranking models for keyphrase search requires sufficient training data. However, previous studies~\citep{bolotova2022non} indicate that the commonly used MSMarco dataset for training neural ranking models primarily consists of question-like queries. In fact, following the \citet{bolotova2022non} strategy, only 0.40\% of the queries are classified as non-questions. Therefore, we propose using generative models to reformulate question-like queries into keyphrase format to address the lack of keyphrase-style datasets.

The contributions of this paper are: (1) We demonstrate the effectiveness of keyphrase-based models in keyphrase search scenarios; (2) we discuss the effect of only training the query encoder versus training both a query and a document encoder; (3) we propose a method to generate keyphrase queries from questions or other query types using Large Language Models (LLMs); (4) we show whether the proposed keyphrase-based training generalises to other models; (5) we manually curate keyphrase queries for the TREC 2019 DL Track test queries to verify the validity of the automatically generated keyphrases; (6) we study the main differences between existing dense retrieval models and our proposals; (7) we assess the performance of both the original model and its keyphrase-enhanced versions within a more realistic, mixed-query scenario; (8) finally, we conduct experiments to evaluate the models' generalisability to different query types, focusing specifically on traditional title-based queries.

Our main finding is that ranking models trained in a dataset primarily composed of keyphrase queries outperform classical dense retrieval approaches in the keyphrase search task while showing comparable performance to original models on question-like, queries. Additionally, we show that the proposed training approach reduces the reliance on lexical matching by enabling the encoding of more relevant information within the special tokens. Finally, our keyphrase-tailored models show improved generalisability compared to the original model, delivering better performance on traditional title-format query collections.

The remainder of this paper is structured as follows: Section~\ref{sec:rw} provides an overview of related work concerning dense retrieval and query and keyphrase generation methods. Our proposed training approach and automatic and manual resource generation process are elaborated upon in Section~\ref{sec:train} and Section~\ref{sec:resources} respectively. The setup and results of our experiments are discussed in Section~\ref{sec:exp}. Finally, we summarise our findings and future work in Section~\ref{sec:concl}.

\section{Related Work}
\label{sec:rw}
Here, we review the state of the art in Pre-trained Language Model (PLM)-based retrieval (\S~\ref{subsec:plm-retrieval}) and existing works in query and keyphrase generation (\S~\ref{subsec:query_kp_gen}). 

\subsection{PLM-based Retrieval} \label{subsec:plm-retrieval}
Cross-encoder models specialise in evaluating sentence or text pair classification. They provide a score for input $\langle query, text \rangle$ pairs, indicating their relationship or similarity.
These models have been successfully used in several tasks, with ranking being one where they have demonstrated exceptional effectiveness.
However, because they require both the query and the document to be processed simultaneously, they are typically employed only in re-ranking scenarios due to their limited efficiency.

Cross-encoders were first introduced by \citet{devlin2019bert} and applied in many NLP tasks over the years.
Regarding retrieval tasks, one of the most well-known examples is \texttt{monoT5}~\citep{nogueira2020document}, a re-ranking model that leverages the text generation capabilities of T5 (Text-to-Text Transfer Transformer)~\citep{raffel2020exploring} to determine whether a document is relevant to a given query. Later, \citet{pradeep2021expando} presented \texttt{duoT5} as part of their re-ranking pipeline, a more robust yet resource-intensive model that evaluates the relevance of two documents for a specific query and selects the most relevant one. Indeed, because of its high computational requirements, \texttt{duoT5} is usually only employed to re-rank the very top (e.g. top-2) documents.

An alternative to overcome the high computational cost of cross-encoders is dense retrieval.
Dense retrieval models leverage contextualised dense representations of both queries and documents to determine relevance scores, allowing both end-to-end and re-ranking strategies. \citet{macdonald2021single} compared the two existing representation strategies used in dense retrieval setups: single representation and multiple representation models.

In single representation approaches, exemplified by \texttt{DPR}~\citep{karpukhin2020dense}, \texttt{ANCE}~\citep{xiong2021approximate}, and \texttt{TCT-ColBERT}~\citep{lin2021batch}, each query or document is encoded into a single embedding. Subsequently, the relevance between the query and document is determined by computing the dot-product of the encoded dense query and document vectors.

However, this study focuses on multiple representation models~\citep{ni2022large, lee2023rethinking, xiao2023reproducibility}, exemplified by  \texttt{ColBERT}~\citep{khattab2020colbert}. The motivation behind this choice mirrors the industry's growing interest in \texttt{ColBERT}~\citep{kim2022applications,clavie2024jacolbertv2,mxbai2024colbert,jha2024jina,intel2024colbert}. Models like \texttt{ColBERT} provide interpretability through late attention mechanisms, identifying the most relevant sections of a document in response to a given query. This is crucial in real-world scenarios where explainability is critical, as it enables users to comprehend the underlying decision-making process. Moreover, it proves invaluable in our research, facilitating the understanding of the subtle differences between question-like and keyphrase-like queries.

Recent work has focused on various fronts, such as distilling knowledge from high-performing models like \texttt{ColBERT} to improve single-representation dense retrieval models~\citep{hofstatter2020improving,hofstatter2021efficiently,lin2020distilling,wang2022inspection}; the impact of negative samples on training~\citep{lin2021batch,zhan2021optimizing}, as selecting proper hard negative examples instead of random negatives improves dense retrieval models performance; embedding pruning and compression to reduce indices size~\citep{lassance2022learned,santhanam2022colbertv2}.
However, to the best of our knowledge, we are the first to study the training and evaluation of dense retrieval models for keyphrase search.

\subsection{Query and Keyphrase Generation} \label{subsec:query_kp_gen}
In this paper, we want to explore the adaptability of dense retrieval models to diverse query types, particularly keyphrase-like queries. For that, we need to create training and testing datasets for keyphrase search. Regarding that objective, we have to remark on the extensive use of generative models in the area.

A well-known example is the emergence of automatic keyphrase generation models, which are replacing traditional extraction-based approaches progressively. Generative models like \texttt{BART}~\citep{kulkarni2022learning}, \texttt{T5}~\citep{gabin2022exploring} and \texttt{GPT}~\citep{song2023chatgpt,martinez2023chatgpt} have proved to outperform existing automatic keyphrase extraction methods, such as \texttt{EmbedRank}~\citep{bennani2018simple} or \texttt{KEA}~\citep{witten1999kea}, in labelling documents with keyphrases. The primary advantage of these models over extractive approaches is their capability to generate keyphrases not explicitly present in the input text.

Another task where generative models have been used recently is query expansion. Until the appearance of these generation-based models, pseudo-relevance feedback (PRF) approaches~\citep{wang2023colbert,wang2023effective} dominated the query expansion task; these leverage documents retrieved by the initial queries to extract terms which are used to expand the queries. Recent studies~\citep{jagerman2023query,wang2023query2doc} leveraged LLMs to generate documents or answers to queries before the ranking phase, aiming to expand the query with them to enhance final rankings.

Query generation techniques have also been used to generate query variants for test collections. \citet{alaofi2023generative} proposed using LLMs to automatically generate query variants from a description of an information need. They employed \texttt{GPT-3.5} to generate variants capable of addressing the information need outlined in the input backstory. Their study reveals a substantial overlap between documents retrieved using human-generated and AI-generated queries.

Generative models have also been used to generate queries to expand documents. For example, \texttt{docT5query}~\citep{nogueira2019doc2query} involves the generation of queries that the input document could potentially answer. These predicted queries are appended to the original documents before indexing. 

Nevertheless, as far as we are aware, no previous research has focused on transforming question-like queries into keyphrase-format queries. We believe this area deserves exploration to facilitate the generation of resources for training models tailored for keyphrase search scenarios, such as academic or professional search~\citep{jacso2015academic,russell2018information,lahiri2024keyphrase}.

\section{Training Keyphrase-based Rankers}
\label{sec:train}
Here, we first explain the functioning of \texttt{ColBERT} (\S~\ref{subsec:prelim}), followed by its adaptation to keyphrase search scenarios (\S~\ref{subsec:proposal}).

\subsection{Preliminaries} \label{subsec:prelim}
In this section, we outline the training process of the \texttt{ColBERT} model, which serves as the backbone model for our approaches.

\looseness -1 First, we can define \texttt{ColBERT} as a linear layer upon the raw token embeddings generated by a Transformer encoder (usually \texttt{BERT}), $\texttt{ColBERT} = Linear((BERT(t_1, ..t_n), m)) \in \mathbb{R}^{m}$, where $m$ is typically set to 128~\citep{khattab2020colbert}. So, the process of producing query ($\phi_q$) and document ($\phi_d$) embeddings using \texttt{ColBERT} can be expressed as:
\begin{align*}
    \phi_q &= \texttt{ColBERT}(\texttt{[CLS]}, \texttt{[Q]}, q_1, ..., q_{|q|}) \in \mathbb{R}^{\mu xm} \\
    \phi_d &= \texttt{ColBERT}(\texttt{[CLS]}, \texttt{[D]}, d_1, ..., d_{|d|}) \in \mathbb{R}^{|d|xm}
\end{align*}
where $\mu$ represents the padded length of the tokenised query.

Then, given a query $q$ and a document $d$, the scoring function for the \texttt{ColBERT} model is defined as:
\begin{equation*}
    score(q,d) = \sum_{i=1}^{|q|}MaxSim(\phi_{q_i}, \phi_d) = \sum_{i=1}^{|q|} \max_{j=1, ...,|d|} Sim(\phi_{q_i}, \phi_{d_j})
\end{equation*}
where $Sim$ represents the function used to compute the similarity between query and document embeddings.

\begin{figure}[t]
    \centering
    \includegraphics[width=0.85\columnwidth]{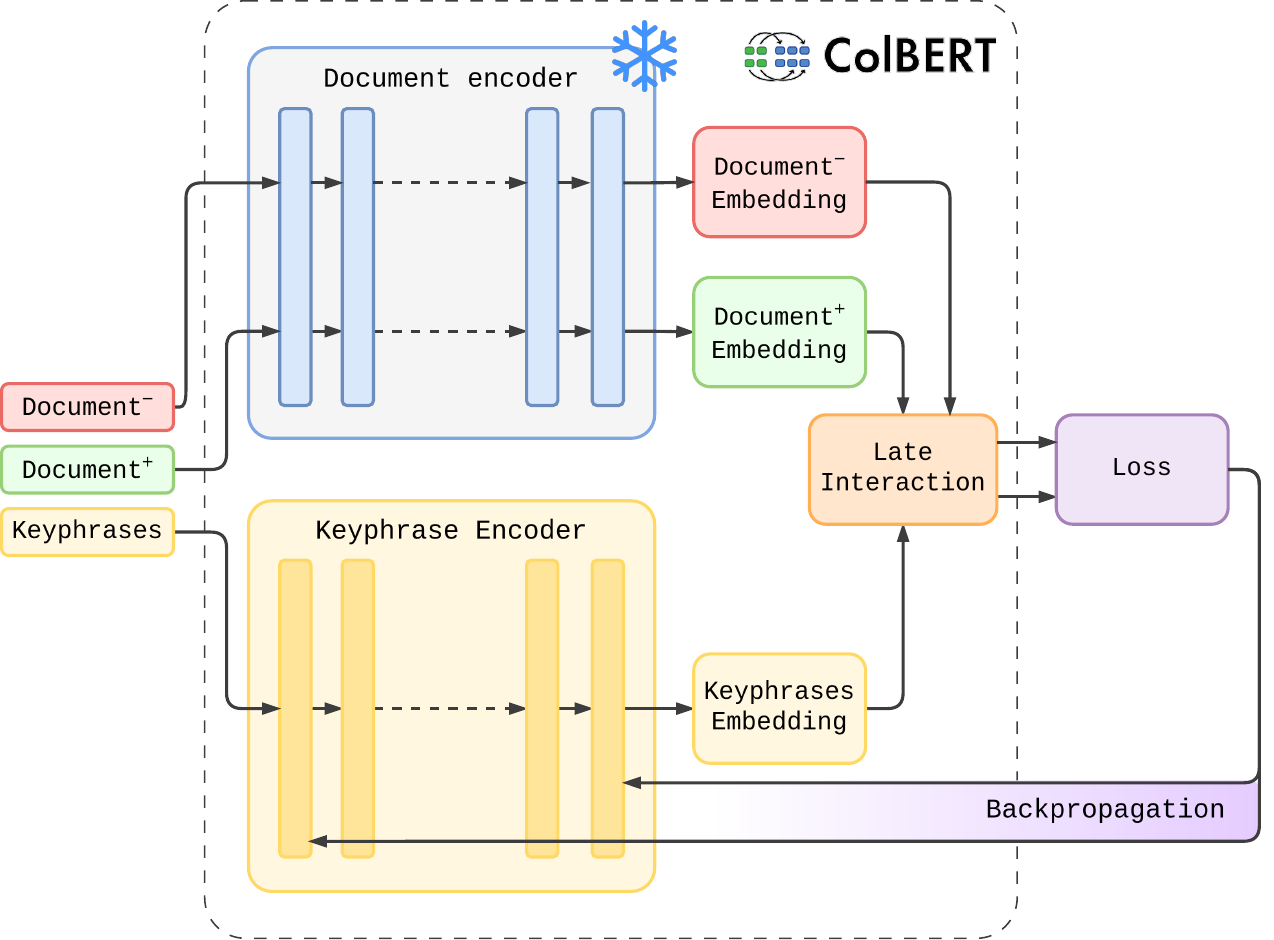}
    \caption{Training process of the \texttt{ColBERTKP$_{Q}$} model using the transformed MSMarco training triples.}
    \label{fig:ColBERTKP$_{Q}$}
    \vspace{-0.5cm}
\end{figure}

The \texttt{ColBERT} model undergoes training using the MSMarco training triples, each triple comprising a query ($q$), a relevant passage ($d^+$), and a non-relevant passage ($d^-$). During training, the objective is to maximise the similarity between the query and relevant passages while minimising similarity with non-relevant passages. 

Therefore, considering a triple $\langle q, d^+, d^- \rangle$, \texttt{ColBERT} computes a score for each document $d^+, d^-$ w.r.t. $q$. This process is optimised using pairwise softmax cross-entropy loss over the computed scores of $d^+$ and $d^-$.

\subsection{From Questions to Keyphrases} \label{subsec:proposal}
Given the example shown in Table~\ref{tab:example}, we argue that training dense retrieval models on datasets like MSMarco tends to bias them towards performing well on question-like queries. 
To address this limitation of current retrieval models, we propose two adaptations to the training process to better suit keyphrase search scenarios.

There are differences in matching behaviour between question-like queries and keyphrase ones that justify the need for training specific models. 
Question-like queries usually contain interrogative adverb phrases (e.g., \textit{"when", "where", "how"}, etc.), which guide the model towards the information need and might match with answer words (e.g. \textit{"why"} with \textit{"because"}).
However, those adverbs are absent in keyphrases, so models must learn to encode that information differently. Keyphrase-tailored models encode this kind of information by using the special tokens (e.g. \texttt{[CLS]}, \texttt{[Q]}, \texttt{[SEP]} and \texttt{[MASK]}) to solve the problem that existing models have.

In the first approach, we train both a keyphrase-like query encoder and a document encoder concurrently, following the standard \texttt{ColBERT} training strategy. This training process optimises the similarity between input keyphrases and relevant documents, while simultaneously minimising similarity with non-relevant ones. By jointly training these encoders, we aim to capture the relationships between keyphrases and documents, thereby improving the effectiveness of the document ranking process in keyphrase search scenarios. We name the resulting model \texttt{ColBERTKP$_{QD}$}.

The second alternative aims to mitigate computational costs associated with training the entire \texttt{ColBERT} model while leveraging its powerful capabilities. In this approach, we freeze the document encoder layer (using the trained weights of a \texttt{ColBERT} checkpoint) while focusing solely on training a keyphrase encoder. We refer to this encoder model as \texttt{ColBERTKP$_{Q}$}.

Figure~\ref{fig:ColBERTKP$_{Q}$} shows the training process of the \texttt{ColBERTKP$_{Q}$} model, where transformed MSMarco triples (\textit{keyphrase query, relevant document, non-relevant document}) are used as inputs. The model generates dense representations for the query and documents, then calculates scores to learn optimal ranking by maximising the score for the relevant document and minimising it for the non-relevant one. Finally, the weights of the keyphrase encoder are updated via backpropagation while the document encoder weights remain constant.

However, training and evaluating a keyphrase-based ranker model necessitates datasets comprising keyphrase queries.
For that,  we transform the queries in the MSMarco training set into keyphrase format.
In Section~\ref{subsec:gen_kps}, we detail how we leverage an LLM to construct datasets specifically tailored for keyphrase search.
Then, instead of the conventional triplet structure, which includes a question, relevant passage, and non-relevant passage, our dataset now features a keyphrase query, along with the corresponding passages.

We formalised our training approaches as follows. 
First, analogously to \texttt{ColBERT}, the process of encoding a keyphrase query $k$ and a document $d$ using \texttt{ColBERTKP$_{QD}$} can be formalised as in Eq.~\ref{eq:colbertkp}. On the other hand, we can define keyphrase query and document embeddings for the \texttt{ColBERTKP$_{Q}$} model as in Eq.~\ref{eq:colbertkpq}.
\begin{equation}
\label{eq:colbertkp}
\begin{aligned}
    \phi_k &= \texttt{ColBERTKP$_{QD}$}(\texttt{[CLS]}, \texttt{[Q]}, k_1, ..., k_{|k|}) \in \mathbb{R}^{\mu xm} \\
    \phi_d &= \texttt{ColBERTKP$_{QD}$}(\texttt{[CLS]}, \texttt{[D]}, d_1, ..., d_{|d|}) \in \mathbb{R}^{|d|xm}
\end{aligned}
\end{equation}

Different to \texttt{ColBERTKP$_{QD}$}, which learns both the query and document encoders, \texttt{ColBERTKP$_{Q}$} uses the original \texttt{ColBERT} model to encode the documents.
\begin{equation}
\label{eq:colbertkpq}
\begin{aligned}
    \phi_k &= \texttt{ColBERTKP$_{Q}$}(\texttt{[CLS]}, \texttt{[Q]}, k_1, ..., k_{|k|}) \in \mathbb{R}^{\mu xm} \\
    \phi_d &= \texttt{ColBERT$_{D}$}(\texttt{[CLS]}, \texttt{[D]}, d_1, ..., d_{|d|}) \in \mathbb{R}^{|d|xm}
\end{aligned}
\end{equation}

Then, the score function for \texttt{ColBERTKP$_{QD}$} and \texttt{ColBERTKP$_{Q}$} given a keyphrase query ($k = \texttt{Q2K}(q)$) and a document ($d$) is:
\begin{equation*}
    score_{kp}(k,d) = \sum_{i=1}^{|k|}MaxSim(\phi_{k_i}, \phi_d) = \sum_{i=1}^{|k|} \max_{j=1, ...,|d|} Sim(\phi_{k_i}, \phi_{d_j})
\end{equation*}

Finally, we define a sequence-to-sequence model as:
\begin{equation*}
    \texttt{Q2K}(seq_{1..N}) \rightarrow seq_{1..M}
\end{equation*}
Specifically, \texttt{Q2K} transforms any query into keyphrase-format, e.g. \texttt{Q2K}(\textit{``how to train a bi-encoder''}) = \textit{``bi-encoder training''}.

Therefore, for a triple $\langle q, d^+, d^- \rangle$ akin to those employed to train \texttt{ColBERT}, we initially convert the query $q$ into a keyphrase-like format using the model \texttt{Q2K}, denoted as $k = \texttt{Q2K}(q)$. Subsequently, for the triple $\langle k, d^+, d^- \rangle$, both \texttt{ColBERTKP$_{QD}$} and \texttt{ColBERTKP$_{Q}$} compute scores individually for each document. Finally, as previously mentioned, the scoring process is optimised using pairwise cross-entropy loss over the scores of $d^+$ and $d^-$.

While the loss computation remains consistent for both the \texttt{ColBERTKP$_{QD}$} and \texttt{ColBERTKP$_{Q}$} models, a disparity appears in their training procedures. In training the former, we optimise weights for both the document and query encoders. In contrast, for the \texttt{ColBERTKP$_{Q}$} model, we exclusively apply backpropagation to the query encoder layers, leaving the document encoder unchanged throughout the training process.

Summing up, we present two training methodologies for dense retrieval models optimised for keyphrase search. These models excel in capturing the subtle relationships between keyphrase-style queries and documents, thereby mitigating the dependence on lexical matching in favour of prioritising special token matching. Moreover, the \texttt{ColBERTKP$_{Q}$} model offers the advantage of compatibility with existing \texttt{ColBERT} indices. This flexibility allows for the utilisation of diverse query encoders based on query type without necessitating multiple document indices.

\section{Resource Alignment for Keyphrase-based Retrieval}
\label{sec:resources}
In this section, we explore the motivation (\S~\ref{subsec:motiv}) behind developing datasets centred on keyphrases and outline the steps taken to construct them. We first show an automated approach (\S~\ref{subsec:gen_kps}) and finish with the creation of a manually curated dataset (\S~\ref{subsec:manual_kps}). Appendix~\ref{appendix:kps_examples} provides examples of both automatically generated and manually annotated keyphrases for selected TREC topics.

\subsection{Motivation} \label{subsec:motiv}
Long-question-format queries serve as the standard for training and evaluating dense retrieval models.
To showcase this fact, in Table~\ref{tab:questions} we show the results of classifying all the MSMarco~\citep{nguyen2016msmarco} dev queries as whether they are questions. Resembling the work by \citet{cambazoglu2021intent} where they present a multi-faceted taxonomy to classify questions asked in web search engines, we use a naive classifier that checks if the query starts with any of the following words: \textit{"what", "when", "where", "which", "who", "whom", "whose", "why", "how", "is", "are", "do", "does"} as a baseline. We also report results using two \texttt{BERT}-based classification models specifically fine-tuned for the task and an LLM instructed to identify whether a query is a question. The accuracy of these methods is assessed against the naive classifier, which may exhibit a few false negatives but is generally conservative in producing false positives.

The results in Table~\ref{tab:questions} reveal that even the less restrictive approach, represented by the naive classifier, categorises over 70\% of the MSMarco queries as questions. Furthermore, both the \texttt{BERT}-based and LLM-based methods exhibit strong agreement with the classifications made by the naive classifier.

\begin{table}[t]
\centering
\caption{Percentage of MSMarco 101k dev queries labelled as questions and accuracy of each method w.r.t. the naive classification approach.}
\renewcommand{\arraystretch}{1.5}
\begin{tabular}{lcc} 
\toprule
\textbf{Method} & \textbf{Questions (\%)} & \textbf{Accuracy w.r.t.\ \texttt{Naive} (\%)} \\
\midrule
\texttt{Naive} & 70.06 & -\\
\texttt{BERT v1}\tablefootnote{\href{https://huggingface.co/shahrukhx01/question-vs-statement-classifier}{shahrukhx01/question-vs-statement-classifier}} & 76.05 & 93.37 \\
\texttt{BERT v2}\tablefootnote{\href{https://huggingface.co/shahrukhx01/bert-mini-finetune-question-detection}{bert-mini-finetune-question-detection}} & 85.77 & 81.36 \\
\texttt{Mistral\tablefootnote{\href{https://huggingface.co/mistralai/Mistral-7B-Instruct-v0.2}{mistralai/Mistral-7B-Instruct-v0.2}}}~\citep{jiang2023mistral} & 81.63 & 75.93 \\
\bottomrule
\end{tabular}
\label{tab:questions}
\end{table}

\subsection{Automatically Generating Keyphrase Queries}
\label{subsec:gen_kps}
Before tackling the training and evaluation of keyphrase-based dense ranking models, it is necessary to have training and testing datasets tailored specifically for keyphrases. However, as previously mentioned, these datasets are scarce. Given this fact, we devise a strategy to transform any query into a keyphrase-format query.

We explore two different approaches for this task. The first strategy employs a T5 model fine-tuned for the automatic keyphrase generation (AKG) task. While keyphrase generation models traditionally process documents to label them with keyphrases, we hypothesise that they may be able to generate keyphrases for queries effectively.

The second approach involves leveraging an LLM specifically instructed to generate keyphrase-format queries given any query. We design the guidelines to direct the LLM's query generation process and complement the prompt with some in-context examples. This prompt can be found in Appendix~\ref{appendix:prompt} and within the code in the repository.

Preliminary experiments reveal that using T5-generated keyphrases as queries leads to a significant decline in the models' ranking performance compared to the original queries. 
Conversely, testing with LLM-generated keyphrases demonstrates comparable performance against the original queries and even some instances of improvement, particularly when employing keyphrase-based ranking models.
The primary difference between these two models lies in the number of generated keyphrases. The T5-based model typically generates a larger set of keyphrases for each query, which aligns with its training process. On the other hand, the LLM-based approach tends to be more concise, often producing either a single keyphrase or a small set thereof.
Given the superior performance of the LLM-based method, we employ it to generate training and test data automatically.

However, we believe that solely relying on LLM-generated data to evaluate the performance of the models is insufficient. First, there is a risk of bias in the evaluation process, as the similarity between the training and testing queries—both generated by the same model—could affect the results. Second, LLM-generated queries may not accurately reflect real-world keyphrase queries.

\subsection{Manually Curating a Test Set}
\label{subsec:manual_kps}
We introduce a manually curated keyphrase-based query set for the TREC DL 2019 dataset to assess the validity of the automatically generated keyphrases. This test set is crafted by three computer science PhDs with experience in the information retrieval field. Their expertise ensures a thorough formulation of queries.

We abstain from providing specific guidelines on query formulation to keep each user's autonomy in query formulation. Users are tasked with generating keyphrase-format queries to address the information needs outlined in the TREC queries. The only directive provided is that they can employ one or multiple keyphrases. We instruct users to separate keyphrases (when using more than one) with commas to adhere to the LLM format.

To facilitate the annotation process, we opt for a straightforward approach. We provide each assessor with a two-column sheet, where one column contains the TREC queries and the other is blank for writing the keyphrase queries. Counting on a manually labelled dataset helps us verify the validity of the automatically generated keyphrase queries and also ensures the reliability of the experiments conducted using them.

\section{Experiments}\label{sec:exp}
We structure our experiments to address the following research questions:
\begin{itemize}
    \item [\textit{RQ1}] \textit{Which keyphrase-based training approach (full or encoder-only) performs better in keyphrase search scenarios?}
    \item [\textit{RQ2}] \textit{Is the performance of keyphrase-based models degraded when using original queries?}
    \item [\textit{RQ3}] \textit{Does the keyphrase-based training generalise to other models?}
    \item [\textit{RQ4}] \textit{Does the effectiveness observed with automatically generated keyphrases extend to manually annotated ones?}
    \item [\textit{RQ5}] \textit{How are keyphrase models and queries different from original queries and models?}
    \item [\textit{RQ6}] \textit{How effectively do the proposed models adapt to mixed query scenarios?}
    \item [\textit{RQ7}] \textit{Does the effectiveness of keyphrase-tailored models translate to traditional keyword (title-only) queries?}
\end{itemize}

\subsection{Experimental Setup}
\label{sec:exp_set}
\subsubsection{Datasets}
We perform experiments on the MSMarco passage corpus, utilising queries from the TREC 2019 and TREC 2020 DL tracks, as well as queries from the MSMarco dev set. Nevertheless, our primary objective is to evaluate the effectiveness of models in keyphrase search scenarios. Therefore, alongside assessing the models' performance on the datasets' original queries, we explore their capability with automatically generated keyphrase-like queries, derived by transforming the original queries into keyphrases. Furthermore, we conduct experiments using manually labelled keyphrases for the TREC 2019 DL queries.

Regarding the training data for the keyphrase-based models, we transform the MSMarco train triples using the LLM-based approach introduced in Section~\ref{subsec:gen_kps}, translating the queries into keyphrase format. Therefore, the dataset we use to train our models is solely formed by keyphrase queries and the corresponding passages.

We also conduct experiments using traditional title-based query collections, employing three query sets—TREC Robust 2004~\citep{vorhees2005robust}, TREC 7~\citep{hawking1998overview}, and TREC 8~\citep{hawking1999overview}—all of which are based on documents from TREC disks 4 and 5.

\subsubsection{Metrics}
We employ commonly used metrics for the TREC 2019 DL and TREC 2020 DL tracks query sets, as outlined in the corresponding track overview papers~\citep{craswell2020overview,craswell2021overview}. Specifically, following common practices, we report the Mean Reciprocal Rank (\texttt{MRR}) and normalised Discounted Cumulative Gain (\texttt{nDCG}) calculated at rank cutoff 10, as well as Mean Average Precision (\texttt{MAP}) at rank cutoff 1000. For the MSMarco dev query set, we report \texttt{MMR@10} and Recall (\texttt{R}) at rank cutoffs 50, 200 and 1000.

For experiments on traditional title-query collections, we report both \texttt{MAP} at rank cutoff 1000 and \texttt{nDCG} at rank cutoff 10.

\subsubsection{Baselines}
In our experiments, we compare the effectiveness of our keyphrase-based models with \texttt{ColBERT}, which serves as the backbone of our work. We use the checkpoint provided by \citet{xiao2023reproducibility} in their reproducibility study, specifically, we employ the cosine similarity version. Additionally, we include results obtained from \texttt{BM25} as a reference. Finally, we examine the performance of one widely recognised re-ranking strategy when experimenting with the generalisation of the proposed training approach: \texttt{monoT5}\footnote{\href{https://huggingface.co/castorini/monot5-base-msmarco}{castorini/monot5-base-msmarco}}.

\subsubsection{Implementation \& Settings}
We conduct all the experiments using PyTerrier~\citep{pyterrier2020ictir} on an Nvidia A100 GPU.

Our training procedure for \texttt{ColBERT}-based models follows the methodology outlined by \citet{xiao2023reproducibility} in their reproducibility study. Beginning with the baseline checkpoint, we train the model with the same number of examples as in~\citep{xiao2023reproducibility}, up to 25k steps with a batch size of 128. Consistently, we employ cosine similarity as our similarity function. Additionally, we experimented with training the keyphrase-based models using \texttt{bert-base} as the initial checkpoint but observed inferior results (not included in the paper for brevity). Regarding parameters configuration, following the original \texttt{ColBERT} paper~\citep{khattab2020colbert}, we set the query padded length ($\mu$) to 32 and the embedding size ($m$) to 128.

For the re-ranking pipelines (indicated by: \texttt{BM25 >> $\mathcal{M}$}), we re-rank the top 1000 documents retrieved by \texttt{BM25} using the model $\mathcal{M}$. 

Regarding statistical significance tests, we compare ranking strategies that belong to the same category and only the keyphrase-format queries: re-rankers with keyphrase queries and end-to-end models with keyphrase queries. By narrowing down our comparisons, we aim to provide more precise insights into the effectiveness of different ranking approaches and models across each search scenario. We use the paired t-test ($p \leq 0.05$) and apply the Holm-Bonferroni multiple testing correction.

For original queries, where our goal is that our proposals achieve equivalent results to existing models rather than surpassing them, we employ the Two One-Sided Test (TOST)~\citep{schuirmann1987comparison}. In this case, we also compare ranking strategies within the same category, using lower and upper bounds of -0.01 and 0.01, respectively, with a significance level set at 0.05.

For the experiments on title-based queries, the documents from TREC disks 4 and 5 contain full text rather than passages. Due to the limitations of BERT-based models, which restrict input to 512 tokens, we segment the full text into passages with a length of 128 and a stride of 64. Then, the score of a document is computed by taking the score of its highest ranked passage, a.k.a., its \textit{max passage}.

Finally, concerning the automatic keyphrase generation process, we employ two distinct approaches for generating the training and test datasets. As detailed in Section~\ref{subsec:gen_kps}, we use an LLM for this task, we specifically leverage an instruction fine-tuned version of \texttt{Mistral}\footnote{\href{https://huggingface.co/mistralai/Mistral-7B-Instruct-v0.2}{mistralai/Mistral-7B-Instruct-v0.2}}~\citep{jiang2023mistral}. For the generation of the training dataset, given the substantial volume of queries to process, we employ an efficient 4-bit precision inference approach~\citep{dettmers2023case}, while original precision is used for building the keyphrase test query sets.

\subsection{Results and Analysis}
\label{sec:results}
In this section, we report and discuss results to answer all the research questions.

\subsubsection{RQ1. Which keyphrase-based training approach (full or encoder-only) performs better in keyphrase search scenarios?} \label{subsec:rq1}
To address our initial research question, we present the ranking performance of both proposed models (\texttt{ColBERTKP$_{QD}$} and \texttt{ColBERTKP$_{Q}$}), using the automatically generated keyphrases for TREC DL and MSMarco dev queries (as presented in Section~\ref{subsec:gen_kps}) and reporting outcomes for both end-to-end ranking and re-ranking strategies.

\begin{table*}[t]
\centering
\setlength{\tabcolsep}{3pt}
\renewcommand{\arraystretch}{1.5}
\caption{Performance of the proposed keyphrase-based models using automatically generated keyphrase queries. $\dagger$ ($\ddagger$) mark denotes significance against the \texttt{ColBERT} re-ranking (end-to-end) baseline.}
\footnotesize
\begin{tabular}{l|lll|lll|llll}
\toprule
\multirow{2}*{\textbf{Model}} & \multicolumn{3}{c|}{\textbf{TREC DL 2019}} & \multicolumn{3}{c|}{\textbf{TREC DL 2020}} & \multicolumn{4}{c}{\textbf{MSMarco Dev Small}} \\
& \texttt{MAP@1k} & \texttt{nDCG@10} & \texttt{MRR@10} & \texttt{MAP@1k} & \texttt{nDCG@10} & \texttt{MRR@10} & \texttt{MRR@10} & \texttt{R@50} & \texttt{R@200} & \texttt{R@1k} \\
\midrule
\texttt{BM25} & 0.2859 & 0.4751 & 0.6120 & 0.2857 & 0.4658 & 0.6416 & 0.1767 & 0.5785 & 0.7311 & 0.8580 \\
\texttt{BM25 >> ColBERT} & 0.4499 & 0.6849 & 0.8516 & 0.4475 & 0.6480 & 0.7585 & 0.3146 & 0.7572 & 0.8324 & 0.8580 \\
\texttt{BM25 >> ColBERTKP$_{QD}$} & \textBF{0.4726}$^\dagger$ & 0.7060 & 0.8876 & 0.4579 & \textBF{0.6881}$^\dagger$ & 0.8076 & 0.3263$^\dagger$ & 0.7685$^\dagger$ & 0.8368$^\dagger$ & 0.8580 \\
\texttt{BM25 >> ColBERTKP$_{Q}$ \& ColBERT$_{D}$} & 0.4688 & \textBF{0.7190}$^\dagger$ & 0.8915 & \textBF{0.4609} & 0.6827$^\dagger$ & 0.7898 & \textBF{0.3300}$^\dagger$ & 0.7654$^\dagger$ & 0.8379$^\dagger$ & 0.8580 \\
\texttt{ColBERT}  & 0.4303 & 0.6869 & 0.8818 & 0.4416 & 0.6348 & 0.7564 & 0.3165 & 0.7795 & 0.8744 & 0.9254 \\
\texttt{ColBERTKP$_{QD}$} & 0.4505 & 0.7072 & 0.8954 & 0.4558 & 0.6831$^\ddagger$ & \textBF{0.8192} & 0.3285$^\ddagger$ & \textBF{0.8017}$^\ddagger$ & \textBF{0.8939}$^\ddagger$ & \textBF{0.9396}$^\ddagger$ \\
\texttt{ColBERTKP$_{Q}$ \& ColBERT$_{D}$} & 0.4573 & \textBF{0.7190} & \textBF{0.9147} & 0.4508 & 0.6728$^\ddagger$ & 0.7970 & 0.3312$^\ddagger$ & 0.7977$^\ddagger$ & 0.8897$^\ddagger$ & 0.9341$^\ddagger$ \\
\bottomrule
\end{tabular}
\label{tab:res_kps}
\end{table*}

Table~\ref{tab:res_kps} reports the performance of the proposed models on automatically generated keyphrase queries. These results show the dominance of keyphrase-based models in the keyphrase search scenario. Both \texttt{ColBERTKP$_{QD}$} and \texttt{ColBERTKP$_{Q}$} outperform the baselines across end-to-end and re-ranking evaluation setups, demonstrating significant differences, particularly evident in the MSMarco Dev Small query set, and, to a smaller extent, in the TREC DL 2019 and TREC DL 2020 datasets.

Upon comparing the performance of our models across both ranking strategies (end-to-end and re-ranking) we observe that while the end-to-end models exhibit slightly better performance, the disparity is not meaningful when considering the cost associated with each strategy. Thus, our approaches demonstrate efficacy across both ranking setups.

Finally, in comparing the fully trained model against the low-resource training approach, which solely trains the query encoder, we see that the encoder-only training not only achieves comparable performance to the fully trained model but also improves its performance in certain setups.

Therefore, in addressing the first research question, we argue that both proposed models outperform existing alternatives in keyphrase search scenarios.
When considering the choice between the fully trained alternative and the encoder-only alternative, the latter is preferable, providing comparable performance with fewer resources required for model training in all evaluation setups.

\begin{tcolorbox}[left*=3mm, right*=3mm, arc=1mm, boxrule=0pt, colback=BoxColor, colframe=white, title=Observation 1, colbacktitle=BoxColorTitle, toptitle=1mm, bottomtitle=1mm]
    Keyphrase-tailored models outperform existing dense retrieval models in keyphrase-based search using both end-to-end and re-ranking strategies.
\end{tcolorbox}


\subsubsection{RQ2. Is the performance of keyphrase-based models degraded when using original queries?}
Having demonstrated the superiority of keyphrase-tailored models in keyphrase search scenarios, we now study their performance when using original (mostly question-like) queries. In this analysis, our objective is not to achieve superior performance for the presented models but to determine their equivalence with existing approaches.

\begin{table*}[t]
\centering
\setlength{\tabcolsep}{3pt}
\renewcommand{\arraystretch}{1.5}
\caption{Performance of the proposed keyphrase-based models using the original query set. $\diamond$ and $\curlywedge$ marks denote statistically equivalence (TOST) against \texttt{ColBERT} re-ranking and end-to-end baselines respectively.} 
\footnotesize
\begin{tabular}{l|lll|lll|llll}
\toprule
\multirow{2}*{\textbf{Model}} & \multicolumn{3}{c|}{\textbf{TREC DL 2019}} & \multicolumn{3}{c|}{\textbf{TREC DL 2020}} & \multicolumn{4}{c}{\textbf{MSMarco Dev Small}} \\
& \texttt{MAP@1k} & \texttt{nDCG@10} & \texttt{MRR@10} & \texttt{MAP@1k} & \texttt{nDCG@10} & \texttt{MRR@10} & \texttt{MRR@10} & \texttt{R@50} & \texttt{R@200} & \texttt{R@1k} \\
\midrule
\texttt{BM25} & 0.3031 & 0.4989 & 0.6780 & 0.2974 & 0.4793 & 0.6457 & 0.1850 & 0.5952 & 0.7481 & 0.8677 \\
\texttt{BM25 >> ColBERT} & 0.4596 & 0.7039 & 0.8353 & \textBF{0.4828} & 0.7146 & 0.8380 & 0.3524 & 0.7909 & 0.8523 & 0.8677 \\
\texttt{BM25 >> ColBERTKP$_{QD}$} & \textBF{0.4608} & 0.7107$^\diamond$ & 0.8469$^\diamond$ & 0.4797$^\diamond$ & 0.7096$^\diamond$ & \textBF{0.8627}$^\diamond$ & 0.3507 & 0.7871 & 0.8524 & 0.8677 \\
\texttt{BM25 >> ColBERTKP$_{Q}$ \& ColBERT$_{D}$} & 0.4579$^\diamond$ & 0.7062$^\diamond$ & 0.8553$^\diamond$ & 0.4816$^\diamond$ & \textBF{0.7154}$^\diamond$ & 0.8356$^\diamond$ & 0.3556 & 0.7841 & 0.8531 & 0.8677 \\
\texttt{ColBERT}  & 0.4452 & 0.7078 & 0.8574 & 0.4692 & 0.6866 & 0.8318 & \textBF{0.3572} & \textBF{0.8221} & \textBF{0.9088} & 0.9491 \\
\texttt{ColBERTKP$_{QD}$} & 0.4378$^\curlywedge$ & \textBF{0.7140}$^\curlywedge$ & \textBF{0.8667}$^\curlywedge$ & 0.4609$^\curlywedge$ & 0.6892$^\curlywedge$ & 0.8576$^\curlywedge$ & 0.3541 & 0.8179 & 0.9081 & \textBF{0.9495} \\
\texttt{ColBERTKP$_{Q}$ \& ColBERT$_{D}$} & 0.4350$^\curlywedge$ & 0.7075$^\curlywedge$ & 0.8566$^\curlywedge$ & 0.4580$^\curlywedge$ & 0.6798$^\curlywedge$ & 0.8602$^\curlywedge$ & 0.3527 & 0.8116$^\curlywedge$ & 0.9040 & 0.9462 \\
\bottomrule
\end{tabular}
\label{tab:res_default}
\end{table*}

Table~\ref{tab:res_default} presents the results of the previous experiment conducted with the original TREC DL and MSMarco queries.
Here, we employ the TOST equivalence test to measure the equivalence between our proposed models and the baselines, as detailed in Section~\ref{sec:exp_set}.
Equivalence tests reveal that keyphrase-based models demonstrate comparable performance on the TREC DL datasets, except for the \texttt{BM25 >> ColBERTKP$_{QD}$} pipeline on TREC DL 2019 in \texttt{MAP@1000}, where results show superiority rather than equivalence.
However, due to the larger query set, equivalence is not consistently observed on the MSMarco dataset. Here, results vary, with both the proposed approaches and the baselines outperforming each other depending on specific scenarios.
These results indicate that the keyphrase-based models achieve comparable performance against the baselines and demonstrate improvements in specific datasets and metrics.

\begin{tcolorbox}[left*=3mm, right*=3mm, arc=1mm, boxrule=0pt, colback=BoxColor, colframe=white, title=Observation 2, colbacktitle=BoxColorTitle, toptitle=1mm, bottomtitle=1mm]
    The performance of keyphrase-based models remains stable across different types of queries, showing equivalent performance to existing dense ranking models.
\end{tcolorbox}

\begin{figure*}[t]
    \centering
    \begin{subfigure}{0.26\textwidth}
         \centering
         \includegraphics[width=\columnwidth]{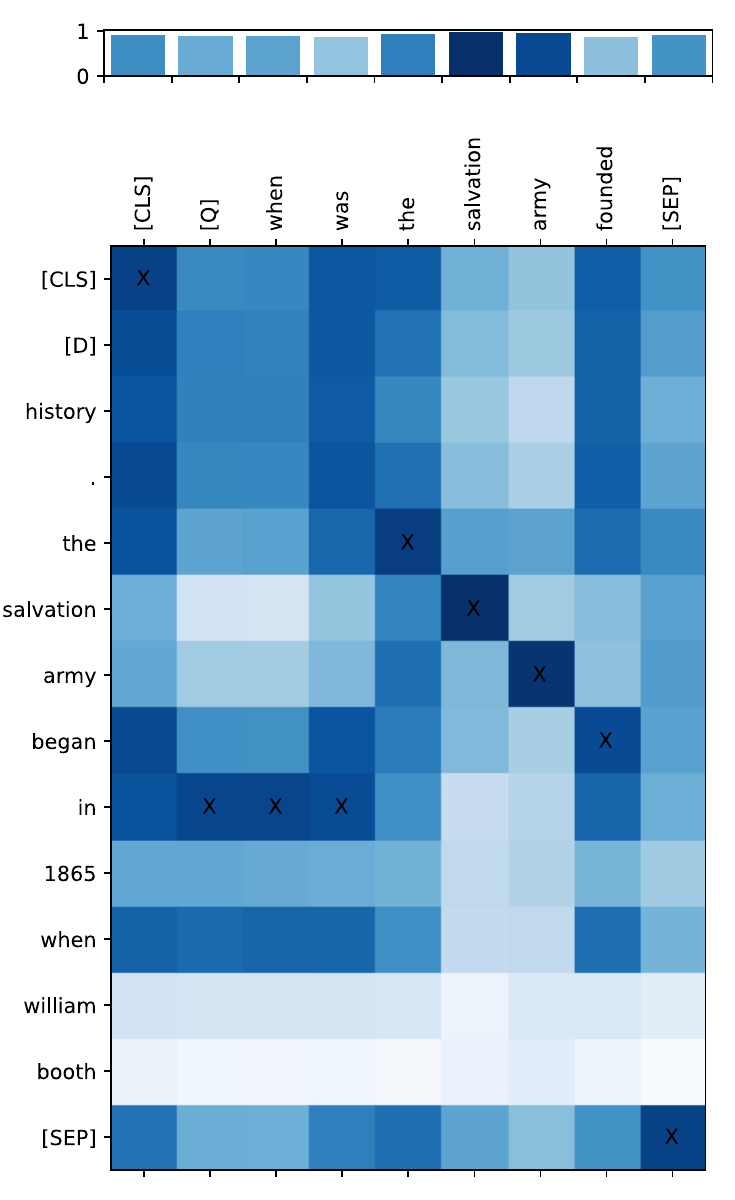}
         \caption{\footnotesize\texttt{ColBERT question}}
         \label{subfig:explain_a}
     \end{subfigure}
    \begin{subfigure}{0.26\textwidth}
         \centering
         \includegraphics[width=\columnwidth]{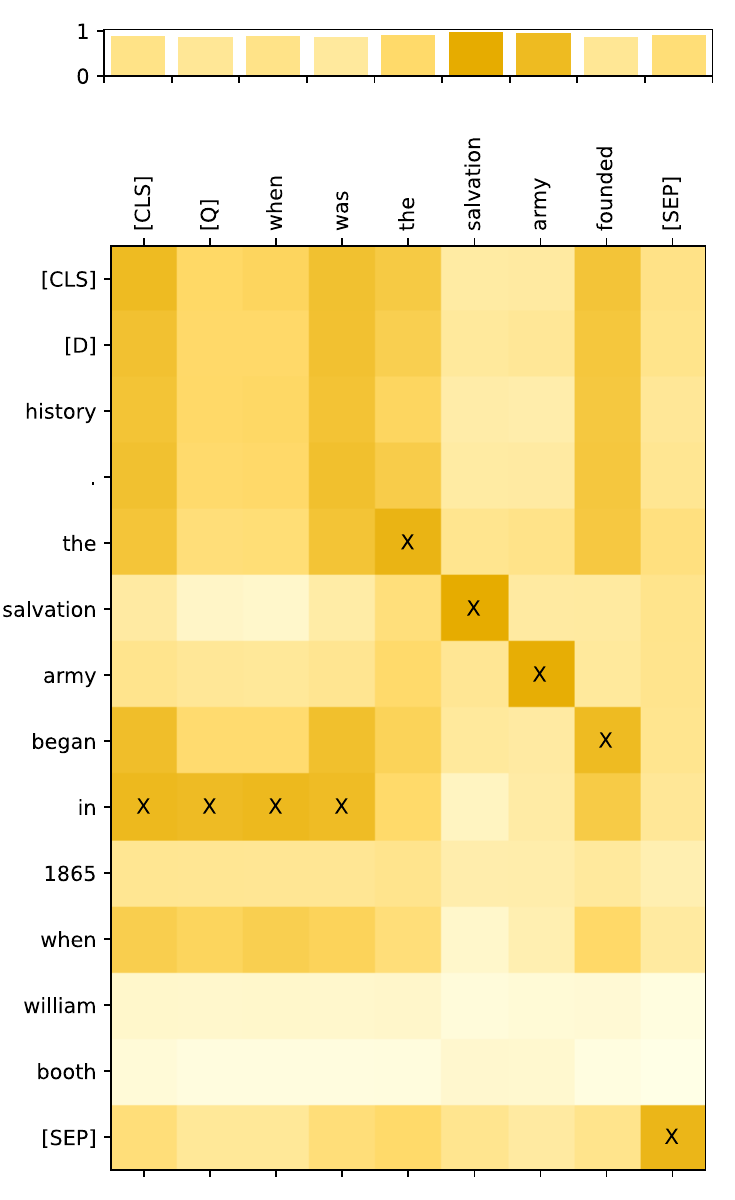}
         \caption{\footnotesize\texttt{ColBERTKP$_{Q}$ question}}
         \label{subfig:explain_b}
     \end{subfigure}
     \begin{subfigure}{0.213\textwidth}
         \centering
         \includegraphics[width=\columnwidth]{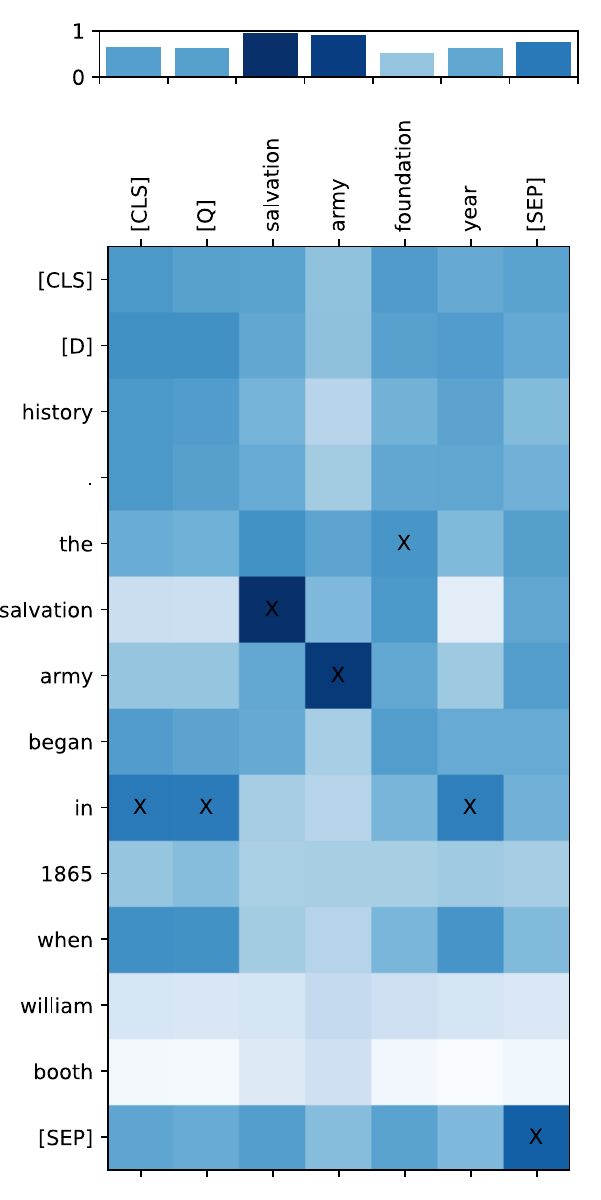}
         \caption{\footnotesize\texttt{ColBERT keyphrase}}
         \label{subfig:explain_c}
     \end{subfigure}
    \begin{subfigure}{0.213\textwidth}
         \centering
         \includegraphics[width=\columnwidth]{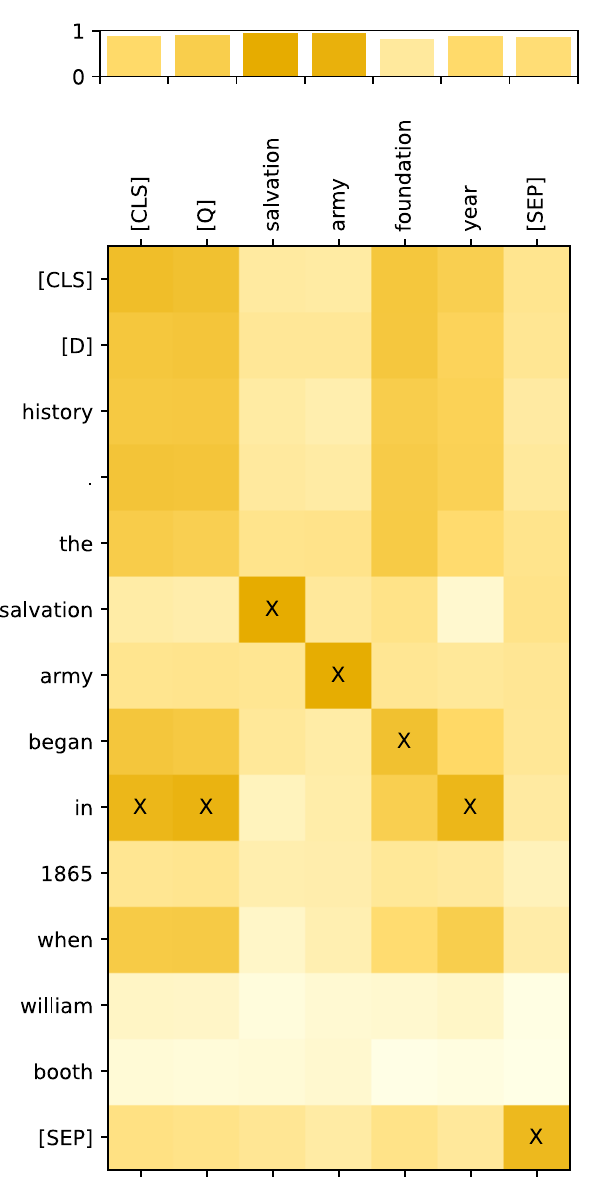}
         \caption{\footnotesize\texttt{ColBERTKP$_{Q}$ keyphrase}}
         \label{subfig:explain_d}
     \end{subfigure}
    \caption{\texttt{ColBERT} and \texttt{ColBERTKP$_{Q}$} interaction for query 962179 (in both original and keyphrase format) and passage 2329699 (shortened). Darker shading in the interaction matrix denotes higher similarity, while the × symbol highlights the document embedding (row) with the highest similarity for each query embedding (column). The histogram above illustrates each query embedding's contribution to the documents's final score, with shading indicating the magnitude of the contribution.}
    \label{fig:explain}
\end{figure*}

Continuing with \textit{RQ1} and \textit{RQ2}, for deeper insights into the superior performance of keyphrase-based models over baselines in keyphrase search scenarios, Figure~\ref{fig:explain} presents interaction diagrams showing the relationship between a query (on both its original and keyphrase versions) and a perfectly relevant document for the query (Table~\ref{tab:example}) for both \texttt{ColBERT} and \texttt{ColBERTKP$_{Q}$} models. Each diagram shows the similarity between query and document embeddings, with darker shading in the interaction matrix indicating higher similarity scores. The × symbol represents the document embedding (rows) with the highest similarity for each query embedding (columns). Additionally, the histogram above each diagram shows the contribution of each query embedding to the document's final score.

Indeed in Table~\ref{tab:res_default}, we observe that the performance between \texttt{ColBERT} and \texttt{ColBERTKP$_{Q}$} is nearly identical when employing original queries, resulting in similar interaction diagrams (Figures~\ref{subfig:explain_a} and \ref{subfig:explain_b}) with minor discrepancies. However, that is not the case for the keyphrase-format query interaction diagrams (Figures~\ref{subfig:explain_c} and \ref{subfig:explain_d}). The similarities between the keyphrase query and the document regarding the \texttt{ColBERT} model greatly diminish, indicating a lower final score for the perfectly relevant document. Conversely, \texttt{ColBERTKP$_{Q}$} exhibits minimal variation from its performance with the original query, yielding high scores and consequently elevating the document's rank.


\subsubsection{RQ3. Does the keyphrase-based training generalise to other models?}

\begin{table}[t]
\setlength{\tabcolsep}{5pt}
\renewcommand{\arraystretch}{1.5}
\caption{Performance of the \texttt{monoT5KP} re-ranking model compared to \texttt{monoT5} and \texttt{ColBERTKP$_{Q}$} re-ranking pipelines. Bold values show the best performance among all setups while underscored values highlight the best results among each query set.}
\begin{tabular}{l@{\hspace{1.5\tabcolsep}}cccc@{\hspace{1.5\tabcolsep}}cccc}
\toprule
\multirow{2}*{\textbf{Model}} && \multicolumn{3}{c}{\textbf{TREC DL 2019}} & &\multicolumn{3}{c}{\textbf{TREC DL 2020}} \\
&& \texttt{MAP@1k} & \texttt{nDCG@10} & \texttt{MRR@10} && \texttt{MAP@1k} & \texttt{nDCG@10} & \texttt{MRR@10} \\
\midrule
\multicolumn{9}{c}{\textit{Original Queries}} \\
\midrule
\texttt{BM25 >> monoT5} && \textBF{0.4815} & \textBF{0.7238} & 0.8798 && \textBF{0.4957} & 0.7138 & \textBF{0.8704} \\
\texttt{BM25 >> monoT5KP} && 0.4597 & 0.7119 & \textBF{0.9019} && 0.4502 & 0.6667 & 0.8201 \\  
\texttt{BM25 >> ColBERTKP$_{Q}$} && 0.4579 & 0.7062 & 0.8553 && 0.4816 & \textBF{0.7154} & 0.8356 \\ 
\midrule
\multicolumn{9}{c}{\textit{Mistral Keyphrases}} \\
\midrule
\texttt{BM25 >> monoT5} && 0.4628 & 0.6982 & 0.7987 && 0.4546 & 0.6801 & 0.8299 \\
\texttt{BM25 >> monoT5KP} && 0.4574 & 0.7060 & \underline{0.8930} && 0.4583 & \underline{0.6920} & \underline{0.8448} \\ 
\texttt{BM25 >> ColBERTKP$_{Q}$} && \underline{0.4688} & \underline{0.7190} & 0.8915 && \underline{0.4609} & 0.6827 & 0.7898 \\ 
\bottomrule
\end{tabular}
\label{tab:monot5}
\end{table}

To study whether training models with keyphrases can be applied beyond the scope of \texttt{ColBERT}, we conducted an experiment that compared the performance of the \texttt{monoT5} baseline from the previous experiment with a \texttt{monoT5} model trained on the same keyphrase-based triples employed in training our proposed \texttt{ColBERT}-based models. We name this model \texttt{monoT5KP}.

We trained the \texttt{monoT5KP} model using the script provided within the PyGaggle~\citep{pradeep2023pygaggle} library. Specifically, we train the model for 5k steps using a batch size of 64 (further training did not improve the model's effectiveness). Different from \texttt{ColBERT}-based approaches, starting from an existing \texttt{monoT5} checkpoint does not result in better performance so we report results using \texttt{t5-base} as the base model.

In Table~\ref{tab:monot5}, we replicate the ranking experiments from \textit{RQ1} and \textit{RQ2} to assess the performance of the \texttt{monoT5KP} model for both original and keyphrase queries (\textit{Mistral Keyphrases}). For brevity, we only report results for the TREC DL datasets. Results show that the keyphrase-tailored model (\texttt{monoT5KP}) exhibits better performance than the original one (\texttt{monoT5}) when assessed with keyphrase queries across all scenarios, except for \texttt{MAP@1k} in the TREC DL 2019 evaluation set. In this case, the difference between both models is subtle, slightly favouring the default training approach.

To further analyse these results, we compare the performance of \texttt{monoT5KP} and \texttt{ColBERTKP$_{Q}$}. For the original queries, both models perform similarly, with \texttt{monoT5KP} showing a slight edge on the TREC DL 2019 dataset and \texttt{ColBERTKP$_{Q}$} slightly ahead on the TREC DL 2020 dataset. When evaluating the keyphrase queries, the two approaches again demonstrate comparable performance: on the TREC DL 2020 dataset, the T5-based model outperforms the ColBERT-based model on two out of three metrics, whereas on the TREC DL 2019 dataset, the ColBERT-based model surpasses the T5-based model in two out of three metrics.

\begin{tcolorbox}[left*=3mm, right*=3mm, arc=1mm, boxrule=0pt, colback=BoxColor, colframe=white, title=Observation 3, colbacktitle=BoxColorTitle, toptitle=1mm, bottomtitle=1mm]
    The keyphrase-based training approach exhibits versatility, as demonstrated by its successful application to models like \texttt{monoT5} whose architecture is a cross-encoder instead of late interaction. This adaptability highlights the applicability of our methodology across diverse model architectures.
\end{tcolorbox}


\subsubsection{RQ4. Does the effectiveness observed with automatically generated keyphrases extend to manually annotated ones?}

\begin{table}[t]
\centering
\setlength{\tabcolsep}{6pt}
\renewcommand{\arraystretch}{1.5}
\caption{Results for manual keyphrases in TREC DL 2019. $\dagger$ ($\ddagger$) mark denotes significance against the \texttt{ColBERT} re-ranking (end-to-end) baseline.}
\begin{tabular}{lccc}
\toprule
\textbf{Model} & \texttt{MAP@1k} & \texttt{nDCG@10} & \texttt{MRR@10} \\
\midrule
\texttt{BM25} & 0.2248 & 0.3893 & 0.5488 \\
\texttt{BM25 >> ColBERT} & 0.3622 & 0.5739 & 0.7093 \\
\texttt{BM25 >> ColBERTKP$_{QD}$} & \textBF{0.3858}$^\dagger$ & 0.5970$^\dagger$ & 0.7416$^\dagger$ \\
\texttt{BM25 >> ColBERTKP$_{Q}$ \& ColBERT$_{D}$} & 0.3816$^\dagger$ & 0.6040$^\dagger$ & 0.7566$^\dagger$ \\
\texttt{ColBERT} & 0.3477 & 0.5717 & 0.7226 \\
\texttt{ColBERTKP$_{QD}$} & 0.3774$^\ddagger$ & 0.6032$^\ddagger$ & 0.7633$^\ddagger$ \\
\texttt{ColBERTKP$_{Q}$ \& ColBERT$_{D}$} & 0.3782$^\ddagger$ & \textBF{0.6131}$^\ddagger$ & \textBF{0.7829}$^\ddagger$ \\
\bottomrule
\end{tabular}
\label{tab:res_manual}
\end{table}

To address this research question, we use the curated test query set established for TREC DL 2019, as outlined in Section~\ref{subsec:manual_kps}. Here, we report metrics using a micro-average strategy instead of computing them individually for each assessor and subsequently averaging their scores.

This experiment aims to validate the conclusions from \textit{RQ1} (\S~\ref{subsec:rq1}). Here, we study if the superior performance of keyphrase-based models when employing automatically generated keyphrase queries extends to keyphrase queries crafted by human assessors.

Table~\ref{tab:res_manual} shows the results of evaluating the proposed models on the TREC DL 2019 dataset using manually annotated keyphrase queries.
For this experiment, we decided to employ only the first keyphrase annotated by each assessor, given that models are usually trained to handle one keyphrase at a time.
This table underscores the dominance of keyphrase-based models over existing dense retrieval models.
Our keyphrase-tailored models outperform all baselines across both end-to-end and re-ranking methodologies, achieving significant improvements on all reported metrics.
This further underscores the importance of developing and deploying ranking models tailored for keyphrase-based search scenarios.

Additionally, a compelling finding from this experiment is that the automatically generated keyphrases outperform those generated by humans, not only for the keyphrase-based models (which are trained using keyphrases generated in the same manner) but also for the baseline models. 

In concluding this experiment, we aim to delineate the primary differences between the assessor-generated keyphrases and those generated by the LLM. 

Table~\ref{tab:stats} shows some statistics about the manually labelled and automatically generated keyphrases.
First, we can see that manual annotators tend to use queries with more keyphrases so that they fully cover the topic they want to search about. In fact,  while the number of keyphrases per query for the automatically generated queries hovers around 1, \textit{Assessor 1} averages nearly 3 keyphrases per query, whereas \textit{Assessor 2} and \textit{Assessor 3} average closer to 1.5 keyphrases per query.
However, the keyphrase length (in terms of words per keyphrase) tends to be slightly shorter for the manually labelled ones.

Finally, we calculate the term overlap between automatic and manually labelled keyphrases defined as:
\begin{equation*}
    O(\mathcal{W}_{auto}, \mathcal{W}_{manual}) = \frac{\left|\mathcal{W}_{auto} \cap \mathcal{W}_{manual}\right|}{min\left(\left|\mathcal{W}_{auto}|, |\mathcal{W}_{manual}\right|\right)}
\end{equation*}
where $\mathcal{W}$ is the set of words for a given keyphrase query.

The results in Table~\ref{tab:stats} reveal a considerable overlap between automatically generated and manually created keyphrases of approximately 75\%. 

\begin{table}[t]
    \centering
    \setlength{\tabcolsep}{6pt}
    \renewcommand{\arraystretch}{1.5}
    \caption{Statistics for manually and automatically generated keyphrases for the TREC DL 2019 test queries.}
    \small
    \begin{tabular}{lcccc}
        \toprule
         & Mistral & Assessor 1 & Assessor 2 & Assessor 3 \\
        \midrule
        \# keyphrases per query & 1.09 & 2.91 & 1.37 & 1.49 \\
        \# words per keyphrase & 3.19 & 2.76 & 2.66 & 3.33 \\
        Overlap & - & 0.72 & 0.73 & 0.80 \\
        \bottomrule
    \end{tabular}
    \label{tab:stats}
\end{table}

\begin{tcolorbox}[left*=3mm, right*=3mm, arc=1mm, boxrule=0pt, colback=BoxColor, colframe=white, title=Observation 4, colbacktitle=BoxColorTitle, toptitle=1mm, bottomtitle=1mm]
    The superior performance of keyphrase-based models using human-generated keyphrase queries reaffirms our findings when utilising automatically generated keyphrases.
\end{tcolorbox}


\subsubsection{RQ5. How are keyphrase models and queries different from original queries and models?}

\begin{figure}
    \centering
    \includegraphics[width=0.80\textwidth]{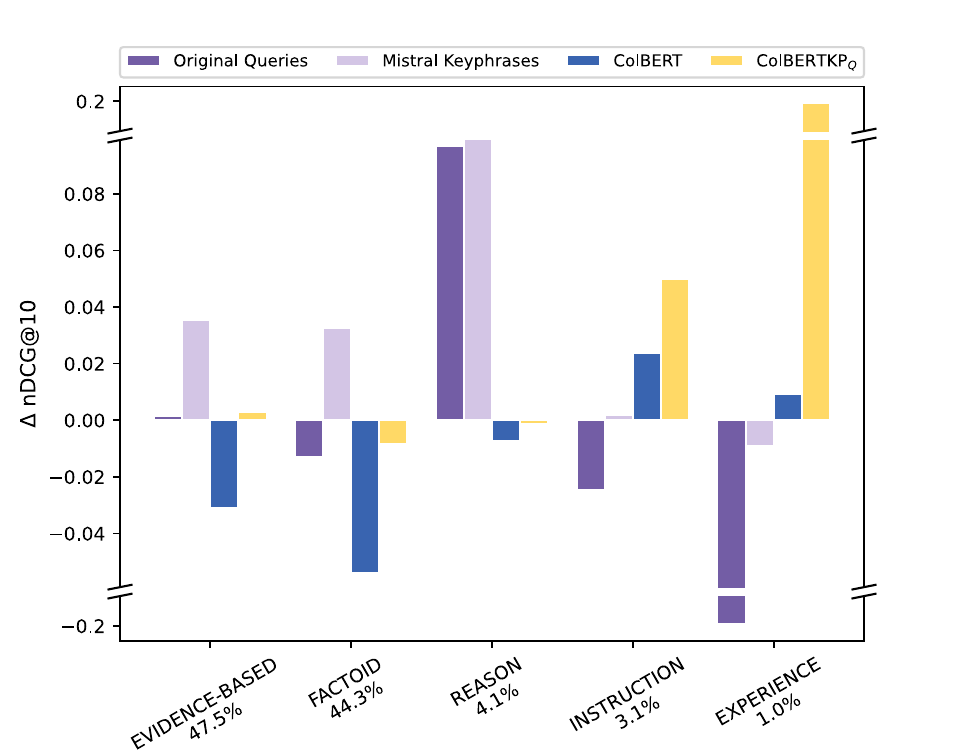}
    \caption{Performance across different types of queries according to the query type taxonomy presented by \citet{bolotova2022non}. Original and Mistral keyphrases bars show the delta between using \texttt{ColBERTKP$_{Q}$} and \texttt{ColBERT} as the retrieval model. \texttt{ColBERTKP$_{Q}$} and \texttt{ColBERT} bars show the delta between using keyphrases and original queries.}
    \label{fig:query_type}
\end{figure}

We want to identify the query types where each model behaves best and show differences in query-document matching behaviour between keyphrase-based and existing models.

Initially, we examine how different model types and queries perform across query types, leveraging the taxonomy proposed in~\citep{bolotova2022non} to categorise queries from the TREC DL 2019 and 2020 datasets. This classification is always done using the original version of the queries. The results of this analysis are shown in Figure~\ref{fig:query_type}.

In Figure~\ref{fig:query_type}, the \textit{Original Queries} and \textit{Mistral Keyphrases} bars illustrate the performance disparity between \texttt{ColBERTKP$_{Q}$} and \texttt{ColBERT}. Positive values mean better performance by the \texttt{ColBERTKP$_{Q}$} model. Here, \texttt{ColBERTKP$_{Q}$} demonstrates resilience in both query types, while \texttt{ColBERT} exhibits a substantial decline when using keyphrase-format queries, particularly in the predominant query types.

Subsequently, the \texttt{ColBERT} and \texttt{ColBERTKP$_{Q}$} bars show the performance variation between automatically generated keyphrases and original queries. Positive values mean better performance by keyphrase queries. We observe that \texttt{ColBERTKP$_{Q}$} outperforms \texttt{ColBERT} across most categories with keyphrase queries while maintaining comparable performance with original queries.

\begin{tcolorbox}[left*=3mm, right*=3mm, arc=1mm, boxrule=0pt, colback=BoxColor, colframe=white, title=Observation 5, colbacktitle=BoxColorTitle, toptitle=1mm, bottomtitle=1mm]
    Keyphrase-based models consistently maintain strong performance across all categories, regardless of whether original or keyphrase-format queries are employed, while the standard model exhibits decreased effectiveness on keyphrase queries.
\end{tcolorbox}

\begin{table}[t]
    \captionof{table}{Impact of different types of matching behaviour for TREC DL 2019 on nDCG@10, and relative decrease from All ($\Delta$). All reported \texttt{ColBERT} results are within the end-to-end dense retrieval scenario.}
    \centering
    \footnotesize
    \renewcommand{\arraystretch}{1.3}
    \setlength{\tabcolsep}{7pt}
    \begin{tabular}{lccccccc} 
    \toprule
    \multirow{2}*{\textbf{Models}} & \textbf{All Types} & \multicolumn{2}{c}{\textbf{Lexical Match}} & \multicolumn{2}{c}{\textbf{Semantic Match}} & \multicolumn{2}{c}{\textbf{Special Match}} \\
    & \texttt{nDCG@10} & \texttt{nDCG@10} & $\Delta$ & \texttt{nDCG@10} & $\Delta$ & \texttt{nDCG@10} & $\Delta$ \\
    \midrule
    \multicolumn{8}{c}{\textit{Original Queries}} \\
    \midrule
    \texttt{BM25} & 0.4795 & - & - & - & - & - & - \\
    \texttt{ColBERT} & 0.7078 & 0.5367 & -24.2\% & 0.0332 & -95.3\% & 0.6820 & -3.6\% \\
    \texttt{ColBERTKP$_{QD}$} & 0.7140 & 0.5053 & -29.2\% & 0.0220 & -96.9\% & 0.6994 & -2.0\% \\
    \texttt{ColBERTKP$_{Q}$ \& \texttt{ColBERT$_D$}} & 0.7075 & 0.5465 & -22.8\% & 0.0375 & -94.7\% & 0.7008 & -0.9\% \\
    \midrule
    \multicolumn{8}{c}{\textit{Mistral Keyphrases}} \\
    \midrule
    \texttt{BM25} & 0.4174 & - & - & - & - & - & - \\
    \texttt{ColBERT} & 0.6869 & 0.5655 & -17.6\% & 0.0178 & -97.4\% & 0.6774 & -1.4\% \\
    \texttt{ColBERTKP$_{QD}$} & 0.7072 & 0.5665 & -19.9\% & 0.0363 & -95.9\% & 0.7030 &  
    -0.6\% \\
    \texttt{ColBERTKP$_{Q}$ \& \texttt{ColBERT$_D$}} & 0.7190 & 0.5706 & -20.6\% & 0.0383 & -94.7\% & 0.7159 & -0.4\% \\
    \bottomrule
    \end{tabular}
    \label{tab:matching}
\end{table}

\begin{wrapfigure}{r}{.30\textwidth}
    \centering
    \includegraphics[width=0.25\textwidth]{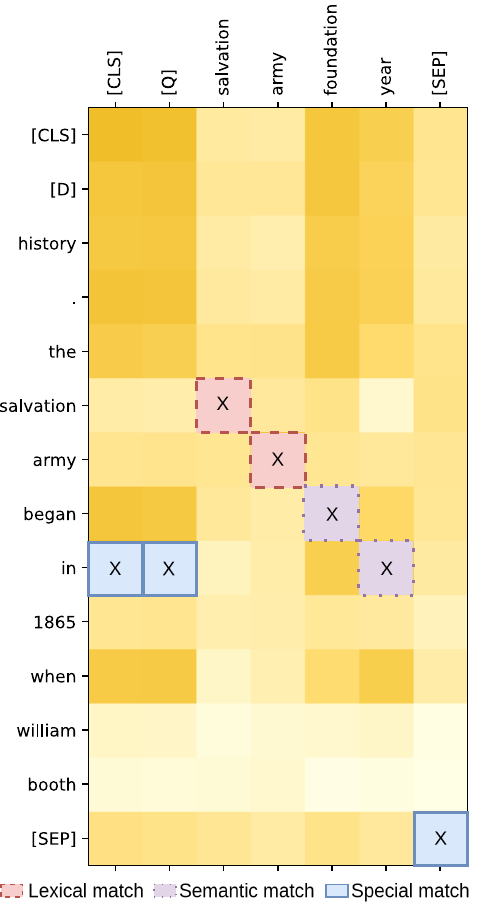}
    \vspace{-0.3cm}
    \caption{Representation of each matching type.}
\end{wrapfigure}
Our second experiment examines how training \texttt{ColBERT} on keyphrases affects its matching behaviour. Following the methodology of \citet{xiao2023reproducibility}, we analyse the three matching strategies between query and document.
\textit{Lexical matching}~\citep{formal2022match} refers to cases where the query and document tokens are exactly the same (e.g., \textit{``salvation''} with \textit{``salvation''}). In contrast, \textit{semantic matching}~\citep{formal2021white} identifies pairs of different tokens (e.g., \textit{``foundation''} with \textit{``began''}), capturing semantic connections between terms. 
Finally, \textit{special token matching} focuses on the alignment of \texttt{ColBERT}'s special tokens, including \texttt{[CLS]}, \texttt{[Q]}, \texttt{[SEP]}, and \texttt{[MASK]}. Using an interaction example from Figure~\ref{fig:explain}, we visually represent each matching type using colours and different line styles in Figure~\ref{fig:query_type}.

Table~\ref{tab:matching} shows retrieval effectiveness through these three matching approaches for the TREC DL 2019 test queries. While both original and keyphrase-based models exhibit similar performance, keyphrase-based models demonstrate an advantage in semantic and special token matching, particularly evident with keyphrase-format queries. This is exemplified in Figure~\ref{fig:explain}, where considerable differences emerge regarding special tokens and semantic matching between \texttt{ColBERT} and \texttt{ColBERTKP$_{Q}$} with the keyphrase-like query. The latter shows higher similarities for the \texttt{[CLS]} and \texttt{[Q]} tokens and for \textit{``foundation''} and \textit{``began''}.

\begin{tcolorbox}[left*=3mm, right*=3mm, arc=1mm, boxrule=0pt, colback=BoxColor, colframe=white, title=Observation 6, colbacktitle=BoxColorTitle, toptitle=1mm, bottomtitle=1mm]
    Training \texttt{ColBERT} with shorter queries enhances special token matching performance, reducing reliance on lexical matches. Token matching nearly achieves parity with the default approach encompassing all matching types.
\end{tcolorbox}


\subsubsection{RQ6. How effectively do the proposed models adapt to mixed query scenarios?}

Building on our analysis of how question-based and keyphrase-based models perform on their respective query types, we now explore how they perform in a mixed scenario where half of the queries are keyphrase-based and the other half are question-like queries.

Specifically, we create a synthetic dataset where 50\% of the queries are question-like (from the original TREC queries), and the remaining are manually annotated keyphrase queries (\S~\ref{subsec:manual_kps}). We compare the performance of the standard \texttt{ColBERT} model with the two proposed approaches (\texttt{ColBERTKP$_{QD}$} and \texttt{ColBERTKP$_{Q}$ \& ColBERT$_{D}$}).

Note that this experiment contains a random component in the query selection process. To mitigate this limitation, we ran the experiment 100 times and averaged the results. In each iteration, 21 queries of each type are randomly selected without overlap. Since three different assessors labelled the manual queries, each run randomly selects queries from one of the assessors.

\begin{table}[t]
\centering
\setlength{\tabcolsep}{6pt}
\renewcommand{\arraystretch}{1.5}
\caption{Results on a hybrid query format scenario using TREC DL 2019 and manually annotated queries.}
\begin{tabular}{lccc}
\toprule
\textBF{Model} & \texttt{MAP@1k} & \texttt{nDCG@10} & \texttt{MRR@10} \\
\midrule
\texttt{BM25 >> ColBERT} & 0.4102 & 0.6373 & 0.7705 \\
\texttt{BM25 >> ColBERTKP$_{QD}$} & 0.4224 & 0.6518 & 0.7908 \\
\texttt{BM25 >> ColBERTKP$_{Q}$ \& ColBERT$_{D}$} & 0.4192 & 0.6530 & 0.8005 \\
\texttt{ColBERT} & 0.4122 & 0.6430 & 0.7769 \\
\texttt{ColBERTKP$_{QD}$} & \textBF{0.4246} & 0.6574 & 0.7971 \\
\texttt{ColBERTKP$_{Q}$ \& ColBERT$_{D}$} & 0.4213 & \textBF{0.6584} & \textBF{0.8067} \\
\bottomrule
\end{tabular}
\label{tab:res_hybrid}
\end{table}

Table~\ref{tab:res_hybrid} presents the retrieval effectiveness of each strategy on the mixed query set. Both \texttt{ColBERTKP$_{QD}$} and \texttt{ColBERTKP$_{Q}$ \& ColBERT$_{D}$} demonstrate comparable performance, outperforming the standard \texttt{ColBERT} model in both re-ranking and end-to-end retrieval setups. When comparing the two proposed approaches, the encoder-only version (\texttt{ColBERTKP$_{Q}$ \& ColBERT$_{D}$}) emerges as the best-performing model.

Table~\ref{tab:res_hybrid} presents the retrieval effectiveness of each strategy on the mixed query set. Both \texttt{ColBERTKP$_{QD}$} and \texttt{ColBERTKP$_{Q}$ \& ColBERT$_{D}$} demonstrate comparable performance, outperforming the standard \texttt{ColBERT} model in both re-ranking and end-to-end retrieval setups. Specifically, the proposed models show gains of up to 3.01\% in \texttt{MAP@1k}, 2.45\% in \texttt{nDCG@10}, and 3.89\% in \texttt{MRR@10}, depending on the setup. When comparing the two proposed approaches, the encoder-only version (\texttt{ColBERTKP$_{Q}$ \& ColBERT$_{D}$}) emerges as the best-performing model.

\begin{tcolorbox}[left*=3mm, right*=3mm, arc=1mm, boxrule=0pt, colback=BoxColor, colframe=white, title=Observation 7, colbacktitle=BoxColorTitle, toptitle=1mm, bottomtitle=1mm]
The mixed query experiment results demonstrate the proposed models' superiority over the standard \texttt{ColBERT} approach. These findings suggest that training the model on both query types can enhance retrieval performance in mixed, real-world, query scenarios.
\end{tcolorbox}


\subsubsection{RQ7. Does the effectiveness of keyphrase-tailored models translate to traditional keyword (title-only) queries?}

\begin{table*}[t]
\centering
\setlength{\tabcolsep}{7pt}
\renewcommand{\arraystretch}{1.5}
\caption{Performance of the proposed keyphrase-based models using traditional title queries. $\dagger$ mark denotes significance against the \texttt{ColBERT} re-ranking baseline.}
\begin{tabular}{lllllll}
\toprule
\multirow{2}*{\textbf{Model}} & \multicolumn{2}{c}{\textbf{TREC Robust 2004}} & \multicolumn{2}{c}{\textbf{TREC 7}} & \multicolumn{2}{c}{\textbf{TREC 8}} \\
& \texttt{MAP@1k} & \texttt{nDCG@10} & \texttt{MAP@1k} & \texttt{nDCG@10} & \texttt{MAP@1k} & \texttt{nDCG@10} \\
\midrule
\texttt{BM25} & 0.2246 & 0.4313 & 0.1719 & 0.4537 & 0.2279 & 0.4764 \\
\texttt{BM25 >> ColBERT} & 0.2490 & 0.4725 & 0.1955 & 0.5023 & 0.2349 & 0.4901 \\
\texttt{BM25 >> ColBERTKP$_{QD}$} & \textBF{0.2612}$^\dagger$ & \textBF{0.4970}$^\dagger$ & \textBF{0.2136}$^\dagger$ & \textBF{0.5595}$^\dagger$ & 0.2417 & \textBF{0.5146} \\
\texttt{BM25 >> ColBERTKP$_{Q}$ \& ColBERT$_{D}$} & 0.2607$^\dagger$ & 0.4914$^\dagger$ & 0.2099$^\dagger$ & 0.5448$^\dagger$ & \textBF{0.2423} & 0.5032 \\
\bottomrule
\end{tabular}
\label{tab:res_title}
\end{table*}

To conclude the experiments section, we examine whether the findings from keyphrase queries also hold for traditional title-query scenarios. We report the ranking performance of both proposed models (\texttt{ColBERTKP$_{QD}$} and \texttt{ColBERTKP$_{Q}$}) on title queries from the TREC Robust 2004, TREC 7, and TREC 8 collections.

Table~\ref{tab:res_title} demonstrates that the effectiveness of the keyphrase models extends to title queries. Both proposed models (\texttt{ColBERTKP$_{QD}$} and \texttt{ColBERTKP$_{Q}$}) outperform the baselines across all datasets, with significant improvements observed on TREC Robust 2004 and TREC 7.

Comparing the fully trained model (\texttt{ColBERTKP$_{QD}$}) with the encoder-only approach (\texttt{ColBERTKP$_{Q}$}) reveals that, while their performances are similar, the former achieves higher scores across nearly all evaluation settings, consistent with previous experiments.

In response to the eighth research question, we argue that training dense retrieval models on a keyphrase-based collection positively impacts performance even when applied to traditional title-based queries.

\begin{tcolorbox}[left*=3mm, right*=3mm, arc=1mm, boxrule=0pt, colback=BoxColor, colframe=white, title=Observation 8, colbacktitle=BoxColorTitle, toptitle=1mm, bottomtitle=1mm]
Adapting models to keyphrase-oriented data enhances their generalisability across different query formats, improving retrieval effectiveness in broader scenarios.
\end{tcolorbox}

\section{Conclusions}
\label{sec:concl}

This paper introduces two keyphrase-based ranking models, namely \texttt{ColBERTKP$_{QD}$} and \texttt{ColBERTKP$_{Q}$}, designed to enhance the performance of dense retrieval systems in keyphrase search scenarios. The former encompasses training the full model architecture, while the latter focuses solely on training the query encoder, reducing training resource requirements. Our experimental setup includes the generation of keyphrase-based datasets for training and testing, adapting original TREC and MSMarco queries using an LLM. Furthermore, we manually label the TREC DL 2019 dataset test queries, providing human-labelled data for performance assessment.

Demonstrating the generalisability of our training approach, we trained a text-to-text re-ranking model (\texttt{monoT5KP}) with promising results. Additionally, we investigate the performance of our models across various query categories, revealing their robustness and effectiveness across different query types.

Our detailed analysis of matching strategies reveals important insights into the behaviour of keyphrase-based models. Training \texttt{ColBERT} with keyphrases enhanced special token matching performance and reduced dependence on lexical matches, demonstrating the adaptability of our approach.

Finally, we demonstrate the adaptability of our approach to other query types, specifically using traditional title-based queries from classic IR collections. Additionally, we conduct tests on hybrid query scenarios containing both keyphrase-like and question-like queries, showing our models' superior performance across these varied formats.

In future work, we plan on testing \texttt{ColBERTv2}~\citep{santhanam2022colbertv2} for keyphrase search. Additionally, applying distillation techniques to refine these models could facilitate more efficient knowledge transfer and model compression without compromising performance. Moreover, creating query and document pruning strategies using keyphrase information could substantially improve the efficiency of the proposed retrieval strategies. Another research involves developing a ranking pipeline that first classifies a query as a question or keyphrase and then selects the appropriate model to encode the query.  Lastly, we aim to explore the application of these models to boolean queries, which use keyphrases as terms joined by boolean operators.

\clearpage
\appendix

\section{Questions to Keyphrases Examples}
\label{appendix:kps_examples}
Table~\ref{tab:example_keyphrases} shows examples of translating original TREC DL 2019 queries into keyphrase format. We report keyphrases generated from three methods: T5 AKG, Mistral, and manual annotations provided by three annotators.

\begin{table}[h]
    \renewcommand{\arraystretch}{1.5}
    \centering
    \footnotesize
    \caption{Examples of automatically generated and manually annotated keyphrases for TREC DL 2019 queries.}
    \label{tab:example_keyphrases}
    \begin{tabular}{p{4cm} p{2cm} p{9cm}}
        \toprule
        \textbf{Query} & \textbf{Method} & \textbf{Generated/Annotated Keyphrases} \\ 
        \midrule
        \multirow{5}{=}{\textbf{do goldfish grow}} & \textbf{T5 AKG} & goldfish, squid \\
        & \textbf{Mistral} & goldfish growth \\
        & \textbf{Annotator 1} & goldfish growth, goldfish size, goldfish max size \\
        & \textbf{Annotator 2} & goldfish size, goldfish max size \\
        & \textbf{Annotator 3} & goldfish growth \\
        \midrule
        \multirow{5}{=}{\textbf{what is wifi vs bluetooth}} & \textbf{T5 AKG} & csm, wifi, bluetooth \\
        & \textbf{Mistral} & wifi vs bluetooth \\
        & \textbf{Annotator 1} & wifi bluetooth advantages, wifi versus bluetooth, wifi vs bluetooth \\
        & \textbf{Annotator 2} & wifi vs bluetooth \\
        & \textbf{Annotator 3} & wifi vs bluetooth, wifi bluetooth differences \\
        \midrule
        \multirow{5}{=}{\textbf{difference between rn and bsn}} & \textbf{T5 AKG} & bsa, rn-bsn, rn \\
        & \textbf{Mistral} & RN vs BSN, registered nurse vs bachelor of science in nursing \\
        & \textbf{Annotator 1} & bachelor science nursing vs registered nurse, bsn vs rn \\
        & \textbf{Annotator 2} & rn vs bsn, bsn rs difference \\
        & \textbf{Annotator 3} & rn vs bsn, rn bsn differences, registered nurse bachelor of science in nursing differences, registered nurse vs bachelor of science in nursing \\
        \midrule
        \multirow{5}{=}{\textbf{rsa definition key}} & \textbf{T5 AKG} & rsa definition, rsa \\
        & \textbf{Mistral} & RSA definition key \\
        & \textbf{Annotator 1} & rsa key, rsa key definition, Rivest Shamir Adleman key \\
        & \textbf{Annotator 2} & rsa definition key \\
        & \textbf{Annotator 3} & rsa key, rsa key definition \\
        \midrule
        \multirow{5}{=}{\textbf{hydrogen is a liquid below what temperature}} & \textbf{T5 AKG} & hydrogen, hydrogen atom \\
        & \textbf{Mistral} & hydrogen liquid temperature \\
        & \textbf{Annotator 1} & hydrogen condensation temperature, liquid hydrogen temperature \\
        & \textbf{Annotator 2} & hydrogen temperature \\
        & \textbf{Annotator 3} & hydrogen melting temperature \\
        \midrule
        \multirow{5}{=}{\textbf{why did the us voluntarily enter ww1}} & \textbf{T5 AKG} & ww1 \\
        & \textbf{Mistral} & us volunteer entry ww1 reason \\
        & \textbf{Annotator 1} & united states word war 1, usa ww1, usa enters ww1, united stated world war I, usa wwI, usa enters wwI \\
        & \textbf{Annotator 2} & us ww1 incentives, us ww1 participation \\
        & \textbf{Annotator 3} & us ww1 voluntary participation causes, us world war 1 voluntary participation causes, united states world war 1 voluntary participation causes, united states ww1 voluntary participation causes \\
    \bottomrule
    \end{tabular}
\end{table}

\clearpage

\section{Prompt Template}
\label{appendix:prompt}

\begin{lstlisting}[caption=Mistral prompt to generate keyphrase-format queries.]
[INST]
You are an expert in transforming a query into keyphrase format. 

You must follow these rules:
1. When the query is a keyphrase you must just return the query and nothing else (examples of keyphrases are: information retrieval, solar panels, quantum computing, heart rate or blockchain).
2. Try not to include variations of the same keyphrase.
3. Acronyms are also considered keyphrases.
4. One-word keyphrases are allowed.
5. Do not provide several sets of keyphrases stick to the output format (do not add "Or: {{<Keyphrases>{{keyphrase 1}}, {{keyphrase 2}}</Keyphrases>}}")

The input format is:
Generate keyphrases for:
<Query>{{input query}}</Query>

The output format is:
<Keyphrases>{{keyphrase 1}}, {{keyphrase 2}}</Keyphrases>

Generate keyphrases for:
<Query>how many calories in jiffy natural peanut butter</Query>
[/INST]
<Keyphrases>jiffy natural peanut butter calories</Keyphrases>
</s>

<s>[INST]
Generate keyphrases for:
<Query>where do belgian draft horses originate</Query>
[/INST]
<Keyphrases>belgian draft horses origin</Keyphrases>
</s>

<s>[INST]
Generate keyphrases for:
<Query>who are the royal guards</Query>
[/INST]
<Keyphrases>royal guard</Keyphrases>
</s>

<s>[INST]
Generate keyphrases for:
<Query>{query}</Query>
[/INST]
\end{lstlisting}

\printcredits

\section*{Acknowledgements}
This work has received support from projects: PLEC2021-007662 (MCIN/AEI/10.13039/501100011033 Ministerio de Ciencia e Innovación, European Union NextGenerationEU/PRTR) and PID2022-137061OB-C21 (MCIN/AEI/10 .13039/501100011033/, Ministerio de Ciencia e Innovación, ERDF A way of making Europe, by the European Union); Consellería de Educación, Universidade e Formación Profesional, Spain (accreditations 2019–2022 ED431G/01 and GPC ED431B 2022/33) and the European Regional Development Fund, which acknowledges the CITIC Research Center. The first author also acknowledges the support of grant DIN2020-011582 financed by the MCIN/AEI/10.13039/501100011033.

\newpage

\end{document}